\documentclass[12pt,a4paper, 5p, twocolumn]{elsarticle}
\usepackage[latin1]{inputenc}
\usepackage{amsmath}
\usepackage{amsfonts}
\usepackage{amssymb}
\usepackage{graphicx}
\usepackage{mathrsfs}
\usepackage{bbold}
\newcommand{\overbar}[1]{\mkern 1.5mu\overline{\mkern-4mu#1\mkern-4mu}\mkern 1.5mu}
\usepackage[]{youngtab}
\journal{Physics Letters B}

\def\Tiny{\fontsize{5pt}{5pt} \selectfont}

\begin{document}

\begin{frontmatter}
\title{Unified theory in the worldline approach \\ DCPT-14/57}

\author{James P. Edwards\fnref{fn}}
\address{Centre for Particle
Theory, Department of Mathematical Sciences, \\
University of Durham, Durham DH1 3LE, UK}
\fntext[fn]{j.p.edwards@dur.ac.uk}

\begin{abstract}
		We explore unified field theories based on the gauge groups $SU(5)$ and $SO(10)$ using the worldline approach for chiral fermions with a Wilson loop coupling to a background gauge field. Representing path ordering and chiral projection operators with functional integrals has previously reproduced the sum over the chiralities and representations of standard model particles in a compact way. This paper shows that for $SU(5)$ the $\bar{\mathbf{5}}$ and $\mathbf{10}$ representations -- into which the Georgi-Glashow model places the left-handed fermionic content of the standard model -- appear naturally and with the familiar chirality. We carry out the same analysis for flipped $SU(5)$ and uncover a link to $SO(10)$ unified theory. We pursue this by exploring the $SO(10)$ theory in the same framework, the less established unified theory based on $SU(6)$ and briefly consider the Pati-Salam model using $SU(4) \!\times \!SU(2)\! \times\! SU(2)$.
\end{abstract}


\end{frontmatter}

\section{Introduction}
The worldline formalism \cite{Strass1, Strass2} is a first quantised approach to field theory and offers a powerful alternative tool for theoretical calculations. Quantities in the field theory are re-expressed as one dimensional quantum mechanical transition amplitudes of spinning point particles. In this context, a recent model of chiral fermions demonstrated an interesting way of summing over the gauge group representations and chiralities present in the standard model \cite{Paul2}. This sum was constructed for a single generation of fermions supplemented by a sterile neutrino. The model is substantially different from the usual field theory approach because the assignment of particles to their group representations and chiralities arises naturally, rather than being pre-determined by hand. The model also has a computational simplicity compared to more traditional methods in field-theory which require the evaluation of a complicated sum over these representations. Instead that sum is generated through the evaluation of a single functional determinant. In this letter we will generalise that result by considering a variety of other symmetry groups that are familiar from previous studies into grand unified theories.

Progress in the worldline description of chiral particles is central to a formulation of the standard model in first quantised language, where the worldline formalism can offer significant computational advantages over calculations in perturbative quantum field theory \cite{Schu, Bast}. Furthermore the first quantised model presented in \cite{Paul2} has an underlying string theory \cite{Us2} which generalises to non-Abelian interactions so it is natural to consider the consequences of using different symmetry groups in that context. 

The motivation for considering alternative gauge groups is the unification of the electroweak and strong interactions. The purpose of this unification is to find a theory with only one coupling constant, from which the standard model emerges after spontaneous symmetry breaking as a low-energy effective theory \cite{Lang}. The gauge group with smallest rank that can accommodate the standard model is $SU(5)$. This is the famous Georgi-Glashow model \cite{Glash}. We shall demonstrate that the representations and chiralities of the standard model particles as described by the standard $SU(5)$ and flipped $SU(5)$ unified theories can also be generated with the new approach of \cite{Paul2}. 

The main results we shall arrive at for the representations and chiralities of standard model particles will be found to agree with well known results in the literature. They can be arrived at by a variety of other group theoretic methods but we believe that the relative compactness of the new approach, combined with the fact that particle multiplets are no longer arbitrarily chosen, means that this approach has some merit as a complementary tool to more conventional methods.

This letter is laid out as follows. The next section briefly reviews the argument and notation in \cite{Paul2} and in Section \ref{secMain} the model is applied to the unified theories of $SU(5)$ and flipped $SU(5)$. We also consider other unified theories which appear in the literature, namely $SU(6)$, $SO(10)$ and $SU(4)\!\times \!SU(2) \!\times \!SU(2)$. 

\section{Fields and worldlines}
We consider a left- or right-handed massless fermion moving in a background gauge field, $A$. We take $A$ to transform in the adjoint representation of some symmetry group which is described by anti-Hermitian Lie algebra generators $\{T_{S}\}$. Working in Euclidean space, the action for a left-handed massless fermion field, $\xi$, is 
\begin{equation}
	S\left[\bar{\xi}, \xi\right] = \int d^{4}x ~i\xi^{\dagger} \bar{\sigma} \cdot D \xi
\end{equation}
where $D = \left(\partial + A\right)$ and $\sigma^{\mu} = \left(\mathbb{1}, \sigma^{i}\right)$ make up the Euclidean Dirac operator $\bar{\sigma}\cdot D$ (the coupling strength is absorbed into $A$). Following the worldline approach requires us to functionally integrate over the matter field to arrive at the effective action $\Gamma\left[A\right]$. In this case, however, we must avoid the well-known problem of how to define the determinant of the Dirac operator acting on chiral fermions transforming in a non-real representation of the gauge group. We can, however, define the phase-difference of determinants which motivates us to consider the variation of the effective action under an infinitesimal change in $A$ \cite{Witten, Mond}. This is easily found to be
\begin{align}
	\delta_{A}\Gamma\left[A\right] &= \delta_{A}\ln{\int \mathscr{D}\!\left(\bar{\xi},\xi\right) e^{-S\left[\bar{\xi}, \xi\right]}}  \nonumber \\
	&=\textrm{Tr}\left(\left(\bar{\sigma}\cdot D\right)^{-1} \bar{\sigma} \cdot \delta A\right)
\end{align}
which can be written in terms of $\gamma$-matrices\footnote{We use $\gamma^{0} = \begin{pmatrix}0 & 1 \\ 1 & 0\end{pmatrix}$ and $\gamma^{j} = \begin{pmatrix}0 & i\sigma^{j} \\ i\bar{\sigma}^{j} & 0\end{pmatrix}$} as
\begin{equation}
	-\int_{0}^{\infty} \! dT \,\textrm{Tr}\left(\frac{\left(1 - \gamma_{5}\right)}{2} e^{T\left(\gamma \cdot D\right)^{2}} \gamma \cdot D\, \gamma \cdot \delta A\right).
	\label{variation}
\end{equation}
We recognise in (\ref{variation}) the heat kernel of the operator $\left(\gamma \cdot D\right)^{2} = D^{2}\mathbb{1} + \frac{1}{2}\gamma^{\mu}F_{\mu\nu}\gamma^{\nu}$ and in \cite{Paul2} a worldline representation of this expression was derived:
\begin{align}
\delta_{A} \Gamma[A] &= -\int_{0}^{\infty}\frac{dT}{T} \oint_{L/R}\!\!\mathscr{D}\omega \mathscr{D}\psi\, e^{-S\left[w, \psi\right]} \nonumber \\
&\qquad\times \mathscr{P} \,\textrm{tr} \left(g\left(2\pi\right) \int_{0}^{2\pi} dt \, \psi \cdot \dot{\omega} \, \psi \cdot \delta A
\right).
\label{world}
\end{align}
Here $\omega^{\mu}\left(t\right)$ describes a point particle traversing a closed loop (which generates the functional trace) and the Grassmann variables $\psi^{\mu}$ are the spin degrees of freedom living on that worldline. The action $S\left[w, \psi\right]$ is just a gauge fixed version of Brink, Di Vecchia and Howe's description \cite{BdVH} of the dynamics of a spin 1/2 point particle:
\begin{equation}
	S\left[\omega, \psi\right] = \frac{1}{2} \int_{0}^{2\pi} \frac{\dot{\omega}^{2}}{T} + \psi \cdot \dot{\psi} \, dt \,,
\end{equation}
from which the integration measure $\frac{dT}{T}$ in (\ref{world}) can be understood as the Faddeev-Popov determinant associated to the fixing of a local worldline supersymmetry.

Upon quantisation the fundamental anti-commutation relations $\left\{\psi^{\mu}, \psi^{\nu}\right\} = \delta^{\mu\nu}$ can be solved by taking $\psi^{\mu} = \frac{1}{\sqrt{2}}\gamma^{\mu}$ which shows that the role of the $\psi^{\mu}$ is to represent the $\gamma$-matrices. The coupling of the fermion to the gauge field is provided by $g\left(t\right)$ -- this is the super-Wilson loop which is familiar from quantum field theory and is often encountered in the worldline approach \cite{Strass1}:
\begin{equation}
	g\left(t\right) = \mathscr{P} \exp{\left(-\int_{0}^{t} \mathcal{A}^{S}(t)T_{S} \,dt\right)}
	\label{gp}
\end{equation}
where
\begin{equation}
	\mathcal{A} = \dot{\omega} \cdot A + \frac{T}{2}\psi^{\mu}F_{\mu\nu}\psi^{\nu}.
\end{equation}
The $L/R$ subscript in (\ref{world}) denotes the boundary conditions on $\psi$ which are interpreted depending on the chirality of the fermion. For left-handed fermions the path integral with periodic boundary conditions on $\psi$ is subtracted from that with anti-periodic boundary conditions whereas for right-handed fermions the two contributions are summed. These combinations insert the appropriate projection operators $1 \mp \gamma^{5}$ into the path integral. For a field theory describing a number of different particles, such as the standard model, one would also need to form the sum of (\ref{world}) over the representations and chiralities of the full matter content. This summation needs to be implemented manually and is determined by the theorist's choice of the assignment of particles into their multiplets.

The path ordering prescription in (\ref{gp}) is required in a non-Abelian theory to ensure gauge invariance of the coupling to the gauge field but it complicates the evaluation of the functional integrals. The conventional way to deal with the non-Abelian nature of the coupling is to perturbatively expand the effective action and to impose the path ordering by hand \cite{Strass2, gluon, Sato}. However, there are other approaches to dealing with the non-commutative character of the Wilson-loop exponent such as by the introduction of additional Grassmann fields \cite{grass}. This was the approach taken in \cite{Paul2} which we now review.

The path ordering can be represented with functional integrals by introducing a set of anti-commuting operators $\tilde{\phi}_{r}$ and $\phi_{s}$ satisfying $\{\tilde{\phi}_{r}, \phi_{s}\} = \delta_{rs}$ with action $S_{\phi} = \int \tilde{\phi} \cdot \dot{\phi}\, dt$ \cite{Sam, Hoker, Ish}. It is easy to check the following definition furnishes us with a representation of the Lie algebra
\begin{equation}
	R^{S} \equiv \tilde{\phi}_{r} T^{S}_{rs}\phi_{s};\qquad \left[R^{S}, R^{T}\right] = if^{STU}R^{U},
	\label{Lie}
\end{equation}
which can be used to absorb the gauge group indices in the Wilson-loop exponent. So instead of working directly with ($\ref{world}$) we will find it advantageous to combine the above ideas to consider as it stands the related quantity
\begin{align}
	\int_{0}^{\infty}\!\frac{dT}{T} \!\int \mathscr{D}\omega \mathscr{D} \psi \,e^{-S\left[w, \psi\right]} \!\int_{0}^{2\pi} \!dt\, \psi \cdot \dot{\omega} \, \psi \cdot \delta A \,\frac{\delta Z\left[\mathcal{A}\right] }{\delta \mathcal{A}} 
	\label{mainP}
\end{align}
where 
\begin{equation}
	Z\left[\mathcal{A}\right] = \int \mathscr{D}\tilde{\phi}\mathscr{D} \phi\,e^{- \int_{0}^{2\pi}\tilde{\phi} \left(\frac{d}{dt} + \mathcal{A}\right)\phi}
	\label{Za}
\end{equation}
is responsible for producing the interaction between the fermion and the gauge field.

This theory has been studied using worldline techniques before \cite{Bastwl1, Bastwl2}, where the focus has been on its canonical quantisation. In particular, the Fock space built by acting on the vacuum with anti-commuting creation operators can be described by wave function components which transform as anti-symmetric tensor products of the representation of the gauge group generators. Acting on wavefunctions of the form $\Psi(x, \tilde{\phi})$ the creation and annihilation operators can be represented by $\phi^{\dagger} = \tilde{\phi}$ and $\phi = \partial_{\tilde{\phi}}$. Then the wavefunctions have a finite Taylor expansion
\begin{equation}
	\Psi(x, \tilde{\phi}) = \Psi(x) + \tilde{\phi}_{r} \Psi^{r}(x) + \tilde{\phi}_{r}\tilde{\phi}_{s} \Psi^{\left[rs\right]}(x) + \ldots,
	\label{Psi}
\end{equation}
where the components transform in fully antisymmetric tensor products of the representation of the Grassmann fields. We shall proceed with functional techniques but will comment on this important property of the theory when we arrive at our results, where the partition function (\ref{Za}) will involve quantities built out of the Wilson-loop in fully anti-symmetric representations. It is possible to extract from (\ref{Za}) the path ordered exponential in (\ref{world}) by projecting onto the sector which transforms in the desired representation \cite{Bastwl1}. However we follow \cite{Paul2}, motivated by the consideration of interacting tensionless spinning strings \cite{Us1, Us2}, by taking advantage of the form of (\ref{Psi}) to generate the super-Wilson loop coupling for multiple particles at once.

Remarkably, in \cite{Paul2} the evaluation of $Z\left[\mathcal{A}\right]$ was shown to provide the correct sum over the chirality and representation assignments for the fermion content of the standard model, augmented by a sterile neutrino. Five pairs of $\tilde{\phi}$ and $\phi$ were used to represent the Lie algebra generators of $SU(3) \! \times\! SU(2) \! \times \! U(1)$. The desired sum of chiralities and representations was found by adding the result of evaluating (\ref{mainP}) with anti-periodic boundary conditions on all Grassmann variables to that with periodic boundary conditions imposed. The information about the representations and chiralities of the particles is contained in the partition function, $Z[\mathcal{A}]$. This novel approach is of great importance for worldline theories of chiral fermions since it greatly simplifies the sum over the representations of the standard model particles.

\section{Unified theory}
\label{secMain}
The assignment used in \cite{Paul2}
\begin{equation}
	\left(T^{S}\right) = i \begin{pmatrix}\frac{1}{2}\lambda^{b} \otimes \mathbb{1}_{2}\\  \mathbb{1}_{3}\otimes \frac{1}{2}\sigma^{a}\\   - \frac{1}{3} \mathbb{1}_{3} \otimes \frac{1} {2} \mathbb{1}_{2}\end{pmatrix}
\end{equation}
incorporating the standard model generators inside $5\!\times\!5$ matrices is reminiscent of the Georgi-Glashow method which embeds the standard model in $SU(5)$. This motivates us to consider this group as the underlying symmetry without purposefully arranging for the standard model content to appear. We shall show that with the direct use of $SU(5)$ as the gauge group the procedure introduced in \cite{Paul2} yields the familiar $\bar{\mathbf{5}}\oplus \mathbf{10} \oplus \mathbf{1}$ representations into which the left-handed matter content of the standard model fits in a manner consistent with the particles' quantum numbers. The chirality associated with these representations will also be in agreement with the Georgi-Glashow model so these well-known assignments are favoured by the new approach.

We must also take five pairs of $\tilde{\phi}$ and $\phi$ to incorporate the generators of $SU(5)$. Then integrating over $\tilde{\phi}$ and $\phi$ in (\ref{Za}) leads to a functional determinant which we define as the product of its eigenvalues. We evaluate this as in \cite{Paul2}:
\begin{align}
	Z\left[\mathcal{A}\right] &= \det{\left(i\left(\frac{d}{dt} + \mathcal{A}\right)\right)} \nonumber \\
	&\propto
	\begin{cases} \det{ \left(\sqrt{g\left(2\pi\right)} + 1/\sqrt{g\left(2\pi\right)}\right)} & \textrm{A/P} \\ 			\det{\left(\sqrt{g\left(2\pi\right)} - 1/\sqrt{g\left(2\pi\right)}\right)} & \textrm{P} \end{cases}
	\label{dets}
\end{align}
where A/P and P refer respectively to anti-periodic and periodic boundary conditions on $\tilde{\phi}$ and $\phi$. Equation (\ref{dets}) shows how the additional Grassmann fields produce quantities related to the Wilson loop $g(2\pi)$ as we proposed earlier.

To calculate the determinants it suffices to name a representation under which the Wilson-loop is to transform. The Lie group valued object $g\left(2\pi\right)$ can then be rotated onto the Cartan subalgebra
\begin{equation}
	g\left(2\pi\right) = \exp{\left(\alpha_{i}H_{i}\right)}\, \qquad i = 1\ldots 4
\end{equation}
whereby its eigenvalue equation can be expressed in terms of the weights of the representation in which it transforms. The goal is then to express the determinants in (\ref{dets}) in terms of group invariant properties of $g\left(2\pi\right)$. In particular, the simplest choice is to take the Wilson loop to transform in the fundamental representation $\mathbf{5}$, whereby we would expect these invariant quantities to be built out of $g(2\pi)$ in the representations made out of fully antisymmetric products of the fundamental (see the discussion of the Fock space associated to the Grassmann fields in the previous section). Indeed we have found that the determinants in (\ref{dets}) can be written as a sum over traces of $g\left(2\pi\right)$ in different representations:
\begin{align}
	&\det{\left(i\left(\frac{d}{dt} + \mathcal{A}\right)\right)} \propto \nonumber \\
	&\mathrm{tr}\left(g_{\mathbf{5}}\right) +  \mathrm{tr}\left(g_{\mathbf{10}}\right) + \mathrm{tr}\left(g_{\mathbf{\overbar{10}}}\right) + \mathrm{tr}\left(g_{\mathbf{\bar{5}}}\right) \nonumber \\
	&+ 2\mathrm{tr}\left(g_{\mathbf{0}}\right)
	\label{5p}
\end{align}
for anti-periodic boundary conditions on $\tilde{\phi}$ and $\phi$ and
\begin{align}
	&\det{\left(i\left(\frac{d}{dt} + \mathcal{A}\right)\right)} \propto \nonumber \\
	&\mathrm{tr}\left(g_{\mathbf{5}}\right) -  \mathrm{tr}\left(g_{\mathbf{10}}\right) + \mathrm{tr}\left(g_{\mathbf{\overbar{10}}}\right) - \mathrm{tr}\left(g_{\mathbf{\bar{5}}}\right)
	\label{5ap}
\end{align}
when periodic boundary conditions are imposed. In the above equations the subscripts denote the representations in which the traces are to be taken and each term describes the Wilson-loop interaction between the gauge field and a particle transforming in the given representation. It is easy to check that these are all of the representations that can be constructed out of anti-symmetric tensor products of the $\mathbf{5}$; in Young Tableaux notation these are\Yvcentermath1{\Tiny$\bullet$}, $\Tiny{ \yng(1)}$, $ \Tiny{ \yng(1,1)}$, $\Tiny{ \yng(1,1,1)}$, $\Tiny{ \yng(1,1,1,1)}$ and $\Tiny{ \yng(1,1,1,1,1)}$.  \Yvcentermath0
\bigskip

These relations can be checked via a quick counting of dimensionality. The determinants in (\ref{dets}) are given by a product of 5 terms, each consisting of combinations of eigenvalues of $g(2\pi)$ in the $\mathbf{5}$ representation (we denote these collectively by $\{\rho_{i}\}$):
\begin{equation}
\det{ \left(\sqrt{g\left(2\pi\right)}\! \pm \!1/\sqrt{g\left(2\pi\right)}\right)} =	\prod_{i = 1}^{5}\left(\rho_{i}^{\frac{1}{2}} \pm \rho_{i}^{-\frac{1}{2}}\right)
\label{count}
\end{equation} 
The product therefore consists of $2^{5}$ terms which must be arranged into invariant quantities associated to the Wilson-loop in various representations. Since the sum of the dimensions of the fully antisymmetric representations of $SU(5)$ is equal to $\sum_{i = 0}^{5} \left(\begin{smallmatrix} 5 \\ i\end{smallmatrix}\right) = (1 + 1)^{5}$, the (signed) combinations of traces given in the preceding equations account for all $32$ of the terms in (\ref{count}).

Following \cite{Paul2}, we correlate the boundary conditions on the spin degrees of freedom $\psi$ with those of $\tilde{\phi}$ and $\phi$, which multiplies the terms in (\ref{5ap}) by a factor of $\gamma_{5}$. The final step is to then take the average of the two contributions in (\ref{5p}) and (\ref{5ap}). The result is the representations and chirality assignments which are well known in $SU(5)$ unified theory:
\begin{align}
	&\left(\mathrm{tr}\left(g_{\mathbf{\bar{5}}}\right) + \mathrm{tr}\left(g_{\mathbf{10}}\right) + 1\right)P_{L}  \nonumber \\ 
	+\,&\left(\mathrm{tr}\left(g_{\mathbf{5}}\right) + \mathrm{tr}\left(g_{\mathbf{\overbar{10}}}\right) + 1\right)P_{R},
	\label{reps}
\end{align}
where $P_{L/R} = \frac{1}{2}(1 \mp \gamma_{5})$ represent the left- and right-handed chirality projection operators respectively. 

The Georgi-Glashow model\footnote{See, for example, \cite{Glash, Bailin}} places a left-handed conjugate down quark colour triplet and isospin singlet and a left-handed isospin doublet (colour singlet) into the $\mathbf{\bar{5}}$ representation. Into the $\mathbf{10}$ representation is placed a left-handed colour triplet and isospin singlet of conjugate up quarks, a left-handed colour triplet and isospin doublet of up and down quarks and a left-handed conjugate electron. It is easy to check that these assignments respect the quantum numbers of the particles if the $\mathbf{10}$ representation is made up out of the anti-symmetric product of two $\mathbf{5}$. The trivial representations that appear in (\ref{reps}) may be relevant to the discussion of neutrino masses, which is important given the evidence for neutrino oscillations. The novel feature of the current approach is that this sum of representations and chiralities was generated by the model once the transformation properties of $g\left(2\pi\right)$ were fixed, rather than needing to be specified by hand. Furthermore, although the Fock space of the $\tilde{\phi}$, $\phi$ theory suggests that $Z[\mathcal{A}]$ should involve objects transforming in these antisymmetric representations, it is notable that these turn out to be the Wilson-loop couplings and that these are associated with the phenomenologically significant chiralities. 

There is another assignment of standard model particles into these representations of $SU(5)$ which appears in the literature. Flipped $SU(5)$ \cite{Barr, Kim} is based on the gauge group $SU(5)\! \times\! U(1)_{X}$. The extra $U(1)$ factor is needed because in this theory the left-handed conjugate \textit{up} quark colour triplet joins a left-handed isospin doublet in the $\mathbf{\bar{5}}$ representation which does not have vanishing weak hypercharge. The $\mathbf{10}$ contains a left-handed colour triplet of conjugate \textit{down} quarks, a left-handed colour triplet and isospin doublet of up and down quarks and a left-handed conjugate \textit{neutrino} with the left-handed conjugate electron now placed in the $\mathbf{1}$ representation. The $SU(5)$ is then broken to $SU(3) \! \times \!SU(2)\! \times\! U(1)_{Z}$ which provides the $SU(3)\!\times\!SU(2)$ part of the standard model; the standard model hypercharge generator is then formed out of a linear combination of $U(1)_{Z}$ and $U(1)_{X}$. 

The simplest way to accommodate an extra $U(1)$ symmetry into our formalism is to include a further generator $T_{25} = \mathbb{1}_{5}$ (this requires us to temporarily suspend the requirement of the tracelessness of the generators arising out of the underlying string model so that the following result could not be used in that context). Then the Wilson loop factorises as
\begin{equation}
	g\left(2\pi\right) = e^{-2i \theta} g_{\mathbf{5}}\left(2\pi\right)\, ; \qquad \theta = \frac{1}{2}\int_{0}^{2\pi}\! \mathcal{A}^{25}\left(t\right) dt.
\end{equation}
It is straightforward to repeat the previous calculation with this extra generator and we find that the terms in (\ref{reps}) simply pick up an extra factor according to their $U(1)$ charge:
\begin{align}
	&\left(\mathrm{tr}\left(g_{\mathbf{\bar{5}}}\right)e^{-3 i \theta} + \mathrm{tr}\left(g_{\mathbf{10}}\right)e^ {i \theta} ~+ e^{5 i \theta} \right)P_{L}  \nonumber \\ 
	+\,&\left(\mathrm{tr}\left(g_{\mathbf{5}}\right)e^{3 i \theta} ~\!~+ \mathrm{tr}\left(g_{\mathbf{\overbar{10}}}\right)e^{- i \theta} + e^{-5 i \theta}\right)P_{R},
	\label{repsflipped}
\end{align}
in agreement with previous assignments used in the literature \cite{Barr}. We shall discuss $SO(10)$ unified theories in more detail below but there is an interesting relation to this group contained in the extra $U(1)$ charges in (\ref{repsflipped}). It is well known that $SU(5)\!\times\! U(1) \subset SO(10)$ \cite{Zee, Masiero} and the $SO(10)$ (spinor) representation $\mathbf{16}$ decomposes as $\mathbf{16} \rightarrow \mathbf{\bar{5}}_{-3} \oplus \mathbf{10}_{1} \oplus \mathbf{1}_{5}$, where the subscripts denote the $U(1)$ charges. This is precisely how the representations associated to the left-handed projection operator have arranged themselves in (\ref{repsflipped}) which is suggestive that it would be natural to further unify the content of this theory into a single $\mathbf{16}$ of $SO(10)$. This is a further point of interest in the current work, where the $U(1)$ charges associated to each multiplet follow only from a specification of the symmetry group, rather than being chosen arbitrarily. 

The multiplets in (\ref{reps}) and (\ref{repsflipped}) are of course well known in the literature, but their generation via the use of the additional Grassmann fields offers a compact way of summing over their contributions to physical phenomena in a first quantised setting. In particular, the worldline approach excels at efficient formulae for multi-loop scattering amplitudes. The current techniques would allow the inclusion of an entire generation of standard model fermions interacting with the $SU(5)$ gauge bosons, without the need to carry out a summation over the matter content by hand, in a generalisation of the approach taken in \citep{Bastwl2}.
\subsection{Other unified theories}
In this subsection we apply the same technique to some other unified theories. Those of interest are those into which the standard model can be embedded and recovered after spontaneous symmetry breaking at some unification scale. The groups $SU(6)$, $SO(10)$ and \linebreak$SU(4) \! \times \!\!SU(2) \!\! \times  \!SU(2)$ feature in the literature and have the property that the $SU(5)$ we have considered above can be embedded into them in a natural way (and so the standard model also fits into these Lie groups). We now determine the representations and chiralities which appear if $g\left(2\pi\right)$ is taken to transform in the fundamental representation of these groups. 

We begin with $SU(6)$, which is not as well established in the literature as the more popular theories based on $SU(5)$. We may ask whether the approach we use here can offer any insight into the use of this group as the underlying symmetry of the field theory by providing a natural assignment of the standard model particles into its representations. We follow the same steps as in the previous section except that we now need six pairs of $\tilde{\phi}$ and $\phi$. Taking $g\left(2\pi\right)$ to transform in the $\mathbf{6}$ representation we find the determinants as follows. For anti-periodic boundary conditions on Grassmann fields
\begin{align}
	&\det{\left(i\left(\frac{d}{dt} + \mathcal{A}\right)\right)} \propto \nonumber \\
	&\mathrm{tr}\left(g_{\mathbf{6}}\right) +  \mathrm{tr}\left(g_{\mathbf{15}}\right) + \mathrm{tr}\left(g_{\mathbf{20}}\right) + \mathrm{tr}\left(g_{\mathbf{\overbar{15}}}\right) + \mathrm{tr}\left(g_{\mathbf{\bar{6}}}\right) \nonumber \\
	&+ 2\mathrm{tr}\left(g_{\mathbf{0}}\right)
\end{align}
and for periodic boundary conditions 
\begin{align}
	&\det{\left(i\left(\frac{d}{dt} + \mathcal{A}\right)\right)} \propto \nonumber \\
	&-\mathrm{tr}\left(g_{\mathbf{6}}\right) +  \mathrm{tr}\left(g_{\mathbf{15}}\right) - \mathrm{tr}\left(g_{\mathbf{20}}\right) + \mathrm{tr}\left(g_{\mathbf{\overbar{15}}}\right) - \mathrm{tr}\left(g_{\mathbf{\bar{6}}}\right) \nonumber \\
	&+ 2\mathrm{tr}\left(g_{\mathbf{0}}\right)
\end{align}
which is multiplied by $\gamma_{5}$ when the boundary conditions on $\psi$ are correlated with those on $\phi$ as described above. Summing the contributions from each set of boundary conditions determines the chiralities selected by the current approach:
\begin{align}
	\left(\mathrm{tr}\left(g_{\mathbf{6}}\right) + \mathrm{tr}\left(g_{\mathbf{20}}\right) + \mathrm{tr}\left(g_{\mathbf{\bar{6}}}\right)\right)&P_{L} \nonumber \\
	+\, \left(\mathrm{tr}\left(g_{\mathbf{15}}\right) + 2 + \mathrm{tr}\left(g_{\mathbf{\overbar{15}}}\right)\right)&P_{R}
	\label{main6}
\end{align}
There have been a few attempts to form a unified theory with gauge group $SU(6)$ \cite{Hart, Fuk}. The general approach places the contents of the $\mathbf{\bar{5}}$ representation of $SU(5)$ into the $\mathbf{\bar{6}}$ of $SU(6)$ along with an exotic fermion, $N$. The $\mathbf{15}$ is constructed as the anti-symmetric product $\mathbf{6} \otimes_{A} \mathbf{6}$ into which fall the remaining standard model particles and conjugate particles to $N$, but in this construction the $\mathbf{\bar{6}}$ and $\mathbf{15}$ multiplets have the same chirality. The result in (\ref{main6}) is inconsistent with this assignment and also suffers from the conjugate representations sharing the same chirality. We have also found a total number of particle representations far exceeding that required for a generation of standard model fields. These problems present a major barrier to phenomenological application of the representations in (\ref{main6}).

We saw in the previous section that the SU(5) and flipped SU(5) theories contained the representations which naturally fit into the $\mathbf{16}$ of SO(10). So for completeness we turn to this gauge group and pursue the link further to investigate whether the results in (\ref{repsflipped}) can be embedded into the larger group. We again take $g\left(2\pi\right)$ to transform in the fundamental representation $\mathbf{10}$ (we use ten pairs of $\tilde{\phi}$ and $\phi$). For anti-periodic boundary conditions on Grassmann fields we evaluate the determinant to be\footnote{Note there is more than one representation of dimension $\mathbf{210}$; to avoid ambiguity the equations refer to that with highest weight $\left[0,0,0,1,1\right]$.}
\begin{align}
&\det{\left(i\left(\frac{d}{dt} + \mathcal{A}\right)\right)} \propto \nonumber \\
	&2\mathrm{tr}\left(g_{\mathbf{10}}\right) + 2 \textrm{tr}\left(g_{\mathbf{45}}\right) + 2\textrm{tr}\left(g_{\mathbf{120}}\right) + \textrm{tr}\left(g_{\mathbf{126}}\right)   \nonumber \\
	& +2\textrm{tr}\left(g_{\mathbf{210}}\right)+\textrm{tr}\left(g_{\mathbf{\overbar{126}}}\right)+  2\textrm{tr}\left(g_{\mathbf{0}}\right).
\end{align}
Similarly for the case of periodic boundary conditions we find
\begin{align}
&\det{\left(i\left(\frac{d}{dt} + \mathcal{A}\right)\right)} \propto \nonumber \\
	&-2\mathrm{tr}\left(g_{\mathbf{10}}\right) + 2 \textrm{tr}\left(g_{\mathbf{45}}\right) - 2\textrm{tr}\left(g_{\mathbf{120}}\right) - \textrm{tr}\left(g_{\mathbf{126}}\right)   \nonumber \\
	&+2\textrm{tr}\left(g_{\mathbf{210}}\right)-\textrm{tr}\left(g_{\mathbf{\overbar{126}}}\right) + 2\textrm{tr}\left(g_{\mathbf{0}}\right)
\end{align}
to which we associate a factor of $\gamma_{5}$. Taking the average of these terms we find the chiralities and representations
\begin{align}
	\left(2\mathrm{tr}\left(g_{\mathbf{10}}\right) + 2\textrm{tr}\left(g_{\mathbf{120}}\right) + \textrm{tr}\left(g_{\mathbf{126}}\right) + \textrm{tr}\left(g_{\overbar{\mathbf{126}}}\right)\right)&P_{L} \nonumber \\
	+ \left(2\textrm{tr}\left(g_{\mathbf{45}}\right)+ 2\textrm{tr}\left(g_{\mathbf{210}}\right) + 2\right)&P_{R}
	\label{main10}
\end{align}
The most common $SO(10)$ model places an entire generation of left-handed standard model particles into the $\mathbf{16}$ representation along with an exotic sterile neutrino \cite{so10, so102} so the assignments we have generated here do not coincide with the well known unified theory. It is unfortunate that the $\mathbf{16}$ representation does not naturally appear out of the approach taken in this work, although it is also not surprising since the current techniques generate multiplets transforming in fully antisymmetric tensor products of the fundamental. In the following section we will consider how we might modify our work to generate this representation. We will also make contact with the Pati-Salam model based on $SU(4)\!\times SU(2)\!\times SU(2)$ (also a subgroup of $SO(10)$).

\subsection{Discussion}
So far we have made the choice to consider the Wilson loop $g\left(2\pi\right)$ transforming in the fundamental representation of each group but this is not necessary and we briefly explore the consequences of alternative representations for this operator. We anticipate higher dimensional representations leading to a larger particle content, since if $g(2\pi)$ has $N$ eigenvalues each determinant consists of $2^{N}$ products. For example, for the case of $SU(3)$, the representations and chiralities were found to be \cite{Paul2}
\begin{align}
	\left( \textrm{tr}\left(g_{\mathbf{\bar{3}}}\right)+1\right)&P_{L} \nonumber \\
	+\left(\textrm{tr}\left(g_{\mathbf{3}}\right)+1\right)&P_{R}
\end{align}
which gave rise to $SU(3)$ triplets and a sterile neutrino. We highlight the sensitivity of this result to the choice of representation for $g\left(2\pi\right)$ by instead taking it to transform in the representation with the next-smallest dimension, $\mathbf{6}$. Then we find the chiralities and representations\footnote{Once again, to avoid ambiguity we note that the $\mathbf{15}$ is the representation with highest weight $\left[2,1\right]$ and its conjugate, $\mathbf{\overbar{15}}$, has highest weight $\left[1,2\right]$. Their Young Tableaux are $\Tiny{ \yng(3,1)}$ and $\Tiny {\yng(3,2)}$ respectively.}
\begin{align}
	Z[\mathcal{A}]_{\mathbf{6}} = \left(\textrm{tr}\left(g_{\mathbf{6}}\right)+\textrm{tr}\left(g_{\mathbf{10}}\right) + \textrm{tr}\left(g_{\mathbf{\overbar{10}}}\right) +\textrm{tr}\left(g_{\mathbf{\bar{6}}}\right)\right)&P_{L} \nonumber \\
	+\left(\textrm{tr}\left(g_{\mathbf{15}}\right)+\textrm{tr}\left(g_{\mathbf{\overbar{15}}}\right) + 2\right)&P_{R}.
	\label{repssix}
\end{align}
These are the anti-symmetric representations constructed out of tensor products of an object whose index takes six values -- these are described by the one column Young Tableaux of $S_{6}$ whose dimensions are in agreement with $Z\left[\mathcal{A}\right]_{\mathbf{6}}$ with the proviso that the $\mathbf{10}$ and $\overbar{\mathbf{10}}$ combine to fill out the $20$ dimensional representation \Yvcentermath1  $\Tiny{ \yng(1,1,1)}$  \Yvcentermath0. The absence of the fundamental representation in (\ref{repssix}) is striking, since this is how the quarks of the standard model transform, so this choice does not seem to be helpful for phenomenological application.

This dependence on the choice of representation for $g\left(2\pi\right)$ motivates us to return to $SU(5)$, where this work has found its best success, and consider the effect of the Wilson loop transforming in the $\mathbf{10}$ representation, which has the next-smallest dimension after the fundamental. We do so to demonstrate how the result depends on the choice of transformation of $g\left(2\pi\right)$. In this case the partition function produces the representations and chiralities
\begin{align}
	\big( \textrm{tr}\left(g_{\mathbf{10}}\right)+ \textrm{tr}\left(g_{\mathbf{50}}\right) + \textrm{tr}\left(g_{\mathbf{70}}\right)   +\textrm{tr}\left(g_{\mathbf{126}}\right) \nonumber \\
	  +  ~\textrm{tr}\left(g_{\mathbf{\overbar{126}}}\right) + \textrm{tr}\left(g_{\mathbf{\overbar{70}}}\right) + \textrm{tr}\left(g_{\mathbf{\overbar{50}}}\right)  + \textrm{tr}\left(g_{\mathbf{\overbar{10}}}\right) \big)&P_{L}  \nonumber \\
	+~\big(\textrm{tr}\left(g_{\mathbf{35}}\right) + \textrm{tr}\left(g_{\mathbf{45}}\right)	+\textrm{tr}\left(g_{\mathbf{175}}\right)+ \textrm{tr}\left(g_{\mathbf{\overbar{175}}}\right)\nonumber \\ +\textrm{tr}\left(g_{\mathbf{\overbar{45}}}\right)+\textrm{tr}\left(g_{\mathbf{\overbar{35}}}\right) + 2\big)&P_{R}.
\end{align}
Note that now the chiralities on each pair of conjugate representations are the same, in contrast to that found when $g\left(2\pi\right)$ was taken to transform in the $\mathbf{5}$ representation. This list of multiplets is obviously not of interest for model building but is included to explain that, although the current approach does not assign particle representations and chiralities by hand, its predictions are very sensitive to the choice of representation for the Wilson loop.

In the previous section we found that the $SU(5)$ and flipped $SU(5)$ theories led to the $\mathbf{\bar{5}} \oplus \mathbf{10} \oplus 1$ representations that can be placed into the $\mathbf{16}$ of $SO(10)$, yet when we considered $SO(10)$ as the underlying gauge group we did not find this representation in our answer. It is worthwhile considering whether we might uncover a connection to the more familiar model by choosing different representations for the Wilson loop. The next smallest representations of $SO(10)$ are the $\mathbf{16}$ and $\overbar{\mathbf{16}}$ and we have calculated the functional determinants for these choices of the representation of $g\left(2\pi\right)$. The chiralities and representations can be expressed in terms of traces which include
\begin{equation}
	\left(\textrm{tr}\left(g_{\mathbf{16}}\right)+ \textrm{tr}\left(g_{\mathbf{\overbar{16}}}\right) + \ldots\right)P_{L}
\end{equation}
but also involve the representations $\mathbf{120}$, $\mathbf{560}$ and further representations whose dimensions exceed 1000. So although the $\mathbf{16}$ can be generated with this choice it brings with it a set of other multiplets not of interest to the building of minimal unified theories. Whilst it would be possible to project onto the $\mathbf{16}$ representation with the introduction of additional $U(1)$ charges \cite{Bastwl1, Bastwl2}, such a manual selection of a certain multiplet is not in keeping with the spirit of the current approach where we restrict our freedom to the choice of representation of the Wilson-loop alone.

We may repeat this analysis also in the case of the Pati-Salam model \cite{PS1} based on the group $SU(4) \! \times \! SU(2)_{L} \! \times SU(2)_{R}$. In this theory the leptons are identified as carrying a fourth colour, extending the $SU(3)$ symmetry of the standard model to $SU(4)$. There is also a second copy of $SU(2)$ which acts on the right-handed particles (which must be broken to yield a low energy theory with only right-handed singlets). Then a single generation of particles can be placed into one left-handed colour quartet and isospin doublet and a second right-handed quartet and isospin doublet \cite{PS1, so102}. This gauge group is also contained in $SO(10)$ and the particle content again fits into a single $\mathbf{16}$. We have considered the simplest case that $g(2\pi)$ transforms under the fundamental representation of this gauge group, finding chiralities and traces which include
\begin{equation}
	\left(\textrm{tr}\left(g_{\mathbf{4}}\right) \textrm{tr}\left(g^{L}_{\mathbf{2}}\right) + \textrm{tr}\left(g_{\mathbf{\bar{4}}}\right) \textrm{tr}\left(g^{R}_{\mathbf{2}}\right) + \ldots \right)P_{L}.
	\label{PS} 
\end{equation}
This contains the $\left(\mathbf{4}, \mathbf{2}, 1\right) \oplus \left(\mathbf{\bar{4}}, 1, \mathbf{2}\right)$ used for the embedding of the standard model: a colour triplet and isospin doublet of left-handed up- and down-quarks joins a colour singlet and isospin doublet of left-handed leptons in the $\left(\mathbf{4}, \mathbf{2}, 1\right)$ whilst their right-handed conjugates transform in the $\left(\mathbf{\bar{4}}, 1, \mathbf{2}\right)$. The total dimension of the representations in (\ref{PS}), however, adds up to 256 so we find a host of unwanted particles alongside those which we sought. Projecting these out would seem unnatural in the context of our current work so (\ref{PS}) does not offer a minimal unified theory.
\section{Concluding remarks}
We have demonstrated that the model in \cite{Paul2} can be used when the symmetry group is $SU(5)$, in which case it provides the low dimensional representations which are used in the Georgi-Glashow model to accommodate the standard model matter content. We chose the Wilson loop to transform in the fundamental representation of $SU(5)$ and discussed how different choices lead to the appearance of different representations and chiralities. The success that we have found for $SU(5)$ and flipped $SU(5)$ suggests that these theories are quite natural in the current framework.

We also considered $SU(6)$ and $SO(10)$ as the gauge groups but for these cases we did not uncover the familiar connections to the standard model. For $SO(10)$ this can only be done by considering the Wilson loop to transform in a higher dimensional representation ($\mathbf{16}$), but this choice also brings unwelcome representations of large dimension. We touched on the $SU(4) \! \times \! SU(2) \! \times \! SU(2)$ theory which suffered the same problem of unwanted representations spoiling the appearance of the familiar particle content. Part of the utility of the approach we have used in this article is the ease with which the gauge group and representation of the Wilson loop can be changed and the consequences of doing so explored.

In combination with the simplifications to some calculations provided by the worldline formulation it would seem valuable to pursue this programme for the standard model and other unified theories. It is certainly simpler to sum the correlated boundary conditions on the functional integrals than to sum over representations and chiralities by hand and the appearance of (\ref{reps}) may provide a guiding principle in how the matter content of the universe can be arranged. It is also useful for calculations using first quantised techniques, where it provides a compact method of summing up the contribution of each of the standard model particles to the scattering of gauge bosons.  

We also hope to relate this to the Abelian string model we have previously used to reformulate QED \cite{Us1, Us2} to extend that theory to non-Abelian gauge symmetry. This will represent useful progress towards reformulating the standard model in terms of worldlines which act as the boundaries of fundamental interacting strings. One aspect we have not touched on here is the introduction of fermion masses through spontaneous symmetry breaking which is an important obstacle to overcome for the development of worldline theories of the standard model.

\section*{Acknowledgements}
It is a privilege to thank Sam Fearn, Henry Maxfield and Paul Mansfield for useful discussions. The author is also grateful to Paul Mansfield and Thomai Tsiftsi for support, patience and for critiquing the manuscript and for some comments from the reviewers. This research is supported by STFC via a studentship and in part by the Marie Curie network GATIS (gatis.desy.eu) of the European Union's Seventh Framework Programme FP7/2007-2013/ under REA Grant Agreement No 317089. \newpage

\bibliographystyle{elsarticle-num-names}
\bibliography{bibSym}
\end{document}